\documentclass[12pt]{article} 
\usepackage{amssymb}
\usepackage{graphicx}
\begin{document} 
\begin{center}
{\large \bf Interplay of total cross sections and ratios of real to imaginary
parts of hadron amplitudes } 

\vspace{0.5cm}                   
{\bf I.M. Dremin}

\vspace{0.5cm}                       
         Lebedev Physical Institute, Moscow 119991, Russia\\

\end{center}

\begin{abstract}
The impact of different assumptions about high energy behavior of the total
cross section of proton-proton interactions on the ratio of the real to 
imaginary part of the forward elastic scattering amplitude is analyzed.
It is shown how experimental data about this ratio at LHC energies can help
in the proper choice of the asymptotic dependence of the total cross section.
\end{abstract}

The total cross sections and the ratios of real to imaginary parts of forward
scattering amplitudes are tightly connected by dispersion relations. Hence, 
both of them should be analyzed simultaneously. We consider the interrelation 
of their behaviors as functions of energy discussing how different assumptions 
about the energy dependence of the total cross section $\sigma _t$ influence 
the predictions of the values of the ratio
\begin{equation}
\rho _0 = \rho (s,t=0) = \frac {{\rm Re}A(s,t=0) }{{\rm Im}A(s,t=0)},
\label{rho}
\end{equation}
where $A(s,t)$ is the elastic scattering amplitude at energy $\sqrt s$ and 
transferred momentum $t$.

In its turn, this ratio measured at LHC can give some hints to the proper
choice of the asymptotic behavior of the total cross section.

Experimentally, this ratio is determined from the interference of the Coulomb
and nuclear parts of the amplitude at extremely small angles. It is negative at
comparatively low energies, becomes positive at energies $\sqrt s$ exceeding
tens of GeV, and increases at ISR. These features are well reproduced by 
dispersion relations.

The total cross sections of hadron processes rise with energy increase. This
surprising fact was first noticed in kaon-proton interactions in Protvino
\cite{prot}. Later, it was confirmed at ISR, S$p\bar p$S, and Tevatron.
Nowadays, with advent of LHC, it has obtained further support.

By the optical theorem, the total cross section is directly related to the
imaginary part of the forward elastic scattering amplitude
\begin{equation}
\sigma _t(s)= \frac {{\rm Im}A(s,t=0) }{s}.
\label{sigt}
\end{equation}     
At the same time, it is well known that due to analyticity properties of the
amplitude its imaginary and real parts are interelated by Kramers-Kronig
integral dispersion relations in such a way that the real part can be expressed
as some integral of the imaginary part, or of the total cross section if the
forward direction is considered. This procedure has been widely used for 
predicting the energy behavior of the real part of the forward scattering
amplitude once the total cross section is inserted under the integral sign
\cite{dnaz, bcahn}. The predictions strongly depend on the assumptions about
the asymptotic behavior of the total cross section. The bunch of predicted 
curves becomes very wide at higher energies. Nevertheless, some qualitative
statements can be done.

The total cross section of proton-proton scattering has been measured in a wide 
energy range. So far, the extrapolations of the total cross sections, which
have been proposed on the basis of various theoretical arguments, cannot be
judged reliable, as can be seen from the fact that many of them have had to be 
discarded as accelerator energies have moved upward. 

At the LHC, the situation simplifies. The energy-decreasing contributions can 
be neglected. Constant term will be mentioned separately. The                                       
phenomenological fits use three main energy-dependent components:
\begin{equation}
\sigma _t(s)= a_1\ln s+a_2\ln ^2s+a_3s^{\Delta}
\label{3com}
\end{equation}     
with variable coefficients $a_i$ and the parameter $\Delta $. The variable $s$
is in GeV$^2$. 

The logarithmic components are related to the so-called
Froissart-Martin bound, which states that the total cross section cannot 
increase asymptotically faster than $\ln ^2s$. Actually, the 
$\ln ^2s$-dependence is often ascribed to the geometric picture of two hadrons
colliding with asymptotically high energies and interacting as Lorentz-contracted
disks with logarithmically increasing radii (for example, see 
\cite{bcahn, ufnd}).

The third component is considered as preasymptotic one, and related to the hard
Pomeron with the intercept $1+\Delta $. We expect that, after unitarization,
it will be consumed at extremely high energies by a slower dependence of the
$\ln ^2s$-type.
 
From Fig. 1 in \cite{dnaz}, we have learned that linear logarithmic increase
of the total cross section leads to very low values of $\rho _0(s)$.
The $\ln ^2s$-dependence saturating the Froissart-Martin bound gives rather
large values of $\rho _0(s)$. In both cases, it slowly decreases with energy
as an inverse logarithm. The power-law dependence provides asymptotically
constant values of $\rho _0$.

These conclusions are supported by the local dispersion relations, which state
that, in practice, the value $\rho _0$
is mainly sensitive to the local derivative of the total cross section. In the
first approximation, the result of the dispersion relation can then be written
in the form \cite{gmig, sukha, fkol}
\begin{equation}
\rho _0(s)\approx \frac {1}{\sigma _t}\left [\tan \left (\frac {\pi }{2}
\frac {d}{d\ln s }\right )\right ]\sigma _t=
\frac {1}{\sigma _t}\left [ \frac {\pi }{2}\frac {d}{d\ln s }+
\frac {1}{3}\left (\frac {\pi }{2}\right )^3\frac {d^3}{d\ln s^3 }+...\right ]
\sigma _t.
\label{rhodi}
\end{equation}
It follows that, at high energies, $\rho _0(s)$ is mainly determined by the 
derivative of the logarithm of the total cross section with respect to the 
logarithm of energy.

According to Eq. (\ref{rhodi}), we obtain
\begin{equation}
\rho _0(s)\approx \frac {\pi }{2}\frac {1}{a_1\ln s+a_2\ln ^2s+a_3s^{\Delta}}
\left [ a_1+2a_2\ln ^2s+a_3\Delta s^{\Delta }
\left (1+\frac {1}{3}(\frac {\pi \Delta }{2})^2\right )\right ].
\label{rhos}
\end{equation}
It is valid at small values of $\Delta $.
                   
First, we consider the three terms of Eq. (\ref{3com}), separately.

1) $a_2=a_3=0. \;\;\;\;\;\;\;$ $\rho _0(s)=\frac {\pi }{2\ln s}$.

It gives the values 0.0887 at energy 7 TeV and 0.0823 at 14 TeV.

2) $a_1=a_3=0. \;\;\;\;\;\;\;$ $\rho _0(s)=\frac {\pi }{\ln s}$.

It gives the values 0.1774 at energy 7 TeV and 0.165 at 14 TeV.

We note that the coefficient in front of $1/\ln s$ is twice larger than in the case 1).

3) $a_1=a_2=0. \;\;\;\;\;\;\;$ 
$\rho _0(s)=\frac {\pi \Delta }{2}\left (1+
\frac {1}{3}(\frac {\pi \Delta }{2})^2\right )$ = const.

It gives the values 0.21 at $\Delta =0.13$ \cite{dnaz}, 0.131 at 
$\Delta =0.08$, and 0.094 at $\Delta =0.06$, i.e., they strongly depend
on the hard Pomeron intercept accepted at the present stage.

These findings are not at all unexpected and comprise with those of Fig. 1
in \cite{dnaz}, but they are somewhat surprising in what concerns the 
predictions at LHC. According to a general folklore 
\cite{dnaz, bcahn, bphi, bhal, kfk}, we would expect this ratio to be in the 
interval 0.13 - 0.14 at LHC energies. The first term in Eq. (\ref{3com}) leads 
to smaller values, and the second term to larger ones. The simple combination of 
them with positive weights does not fit either the total cross sections or the
expected values of $\rho _0$. Sometimes, the coefficient $a_1$ is chosen to 
be negative (see, e.g., \cite{bhal, gross}). In these cases, the constant term
should be added in the expression for the total cross section. 

The old version
\cite{gross} predicts too high value of 127 mb for the total cross section at
7 TeV. It has been used in extrapolation 4 of 
\cite{dnaz}, and gave rise to a maximum and then to somewhat faster 
decrease of $\rho _0$ with energy but its values are still a little bit large. 
Surely, this is related to larger cross sections predicted at LHC.

The recent fit in \cite{bhal} predicts slightly smaller value 95.4$\pm $1.1 mb
compared with 98.3$\pm $2.9 mb in experiment but acceptable in view of error
estimates. The values of $\rho _0$ are within the interval 0.13 - 0.14.

However, up to now we have no reasonable justification for including the terms 
with negative sign in the expressions for total cross sections. Actually,
the total cross section is compiled as a sum of positive
contributions of individual channels. Hence, one should develop a model with
such negative contributions in particular channels. 

The intermediate behavior between variants 1) and 2) was considered in
\cite{grau, grau1}. It is supposed that $\sigma _t \propto \ln ^rs$ with 
$1\leq r\leq 2$.
Then $\rho _0(s)=\pi r/2\ln s$, and fills in the gap between predictions of
1) and 2). However, the value of $r$ is left indefinite. It is claimed that 
according to some fits of experimental data it can be within the limits
$1.3\leq r\leq 1.52$. It would mean $0.115\leq \rho _0\leq 0.135$ at 7 TeV,
and $0.107\leq \rho _0\leq 0.125$ at 14 TeV. The total cross section would
increase, correspondingly, by 10 - 12$\%$.

It looks as if the 
admission of the third term is necessary not only for fits of the total cross
section but is required by the values of $\rho _0$. Its prediction of constant
$\rho _0$ for a constant $\Delta $ is, however, not very appealing intuitively.
We would expect that the parameter $\Delta $ depends on energy, and, probably,
becomes smaller at higher energies to prevent the violation of the 
Froissart-Martin bound. Theoretically, it would correspond to unitarization
of the hard Pomeron. Then, $\rho _0(s)$ becomes smaller than 0.13 as
demonstrated above for $\Delta =0.06$.

In general, lower values of $\rho _0$ at LHC compared to the widely 
accepted interval 0.13 - 0.14 would indicate smaller parameter
$\Delta $, but show that the way to asymptopia is still very long.
Larger values seem less probable, but, if observed, would imply more 
intriguing situation with competition between pure $\ln ^2s$-law and larger
values of the hard Pomeron intercept $\Delta $.

There were attempts to keep the total cross section within $\ln ^2s$-bound
suppressing it by small power of $s$ 
and fit $\rho _0(s)$. For example, such a shape of the total cross section
was proposed in \cite{okor} with the slight power-like damping
\begin{equation}
\sigma _t \propto cs^{-\alpha}\ln ^2s
\label{oko}
\end{equation}
with $\alpha \approx 0.02$.

It leads to
\begin{equation}
\rho _0(s)=\frac {\pi }{\ln s}(1-\frac {\alpha }{2}\ln s).
\label{rhook}
\end{equation}
This results in extremely low value at 7 TeV $\rho _0(7 TeV)=3\cdot 10^{-4}$,
and even negative value at 14 TeV $\rho _0(14 TeV)=-2.6\cdot 10^{-2}$, which
look unrealistic. Nevertheless, this teaches us that slight variations of 
$\ln ^2s$-law can drastically diminish $\rho _0(s)$.

The above results can be considered as upper limits of the corresponding 
estimates for particular models. In most models the energy independent term
$\sigma _0$ is added in Eq. (\ref{3com}). Therefore it appears in the
denominator of Eq. (\ref{rhos}) and diminishes the above estimates of $\rho _0$.
It is especially large in the models \cite{bcahn, gross, grau}, where it exceeds
even the minimum of the cross section at tens of GeV, comparatively
small in \cite{kfk, okor, ama1, ama2}, and is absent in \cite{lipk}.

To conclude, we have argued that neither of the three possibilities of energy
behavior of the total cross section presented in Eq. (\ref{3com}) seem to
satisfy separately our experience with energy dependences of $\sigma _t(s)$
and $\rho _0(s)$. Only their combination with some constant term added seems 
to be satisfactory. Which one of
these contributions wins in the competition for asymptopia, can be guessed
from the forthcoming data on $\rho _0$ at LHC energies.  

This work was supported by RFBR grant 12-02-91504-CERN-a and
by the RAN-CERN program.

\end{document}